\documentclass[12pt,preprint]{aastex}  
\usepackage{graphicx}
\usepackage{txfonts}
\newcommand{\va}{v_{\mathrm{A}}}
\newcommand{\cs}{c_{s}}
\newcommand{\ct}{c_{\mathrm{T}}}

\newcommand{\vap}{v_{\mathrm{A}f}}
\newcommand{\csp}{c_{\mathrm{s}f}}
\newcommand{\ctp}{c_{\mathrm{T}f}}

\newcommand{\pd}{\partial}

\begin{document}

	\title{MAGNETOHYDRODYNAMIC WAVES IN A PARTIALLY IONIZED FILAMENT THREAD}

	\shorttitle{MAGNETOHYDRODYNAMIC WAVES IN A PARTIALLY IONIZED FILAMENT THREAD}

   \author{R. Soler, R. Oliver, and J. L. Ballester}
   \affil{Departament de F\'isica, Universitat de les Illes Balears,
              E-07122, Palma de Mallorca, Spain}
              \email{[roberto.soler;ramon.oliver;joseluis.ballester]@uib.es}

  \begin{abstract}

Oscillations and propagating waves are commonly seen in high-resolution observations of filament threads, i.e., the fine-structures of solar filaments/prominences. Since the temperature of prominences is typically of the order of $10^4$~K, the prominence plasma is only partially ionized. In this paper, we study the effect of neutrals on the wave propagation in a filament thread modeled as a partially ionized homogeneous magnetic flux tube embedded in an homogeneous and fully ionized coronal plasma. Ohmic and ambipolar magnetic diffusion are considered in the basic resistive MHD equations. We numerically compute the eigenfrequencies of kink, slow, and Alfv\'en linear MHD modes, and obtain analytical approximations in some cases. We find that the existence of propagating modes is constrained by the presence of critical values of the longitudinal wavenumber. In particular, the lower and upper frequency cut-offs of kink and Alfv\'en waves owe their existence to magnetic diffusion parallel and perpendicular to magnetic field lines, respectively. The slow mode only has a lower frequency cut-off, which is caused by perpendicular magnetic diffusion and is significantly affected by the ionization degree. In addition, ion-neutral collisions is the most efficient damping mechanism for short wavelengths while ohmic diffusion dominates in the long-wavelength regime. 

  \end{abstract}



\section{INTRODUCTION}

Solar prominences and filaments are large-scale magnetic structures located in the solar corona. They are primarily composed of hydrogen ($\sim$90\%) with a much smaller quantity of helium ($\sim$10\%), their composition being similar to solar and cosmic abundances. The physical properties of their plasma are very different from those typical of the coronal medium. The prominence temperature is of the order of $10^4$~K, i.e., a hundred times smaller than the coronal temperature, whereas the prominence density is around two orders of magnitude larger than that of the surrounding coronal plasma. For these reasons, the very existence of cold and dense filaments embedded in such an ``extreme'' coronal environment is not well-understood yet, but certainly the coronal magnetic field must play a crucial role in the formation, support against gravity, and thermal isolation of prominences. Due to their relatively cool temperature, the prominence plasma is only partially ionised. However, the exact ionization degree of prominences is unknown and the reported ratio of electron density to neutral hydrogen density \citep[e.g.,][]{patsu} covers about two orders of magnitude (0.1 -- 10). While the support of ions and electrons against gravity can be explained by the magnetic force, the support of neutral atoms seems more problematic because they are insensible to the magnetic field. Regarding this problem, a plausible explanation for the support of neutrals is by means of collisions with ions \citep{pecseli,gilbert02,gilbert07}.

High-resolution H$\alpha$ observations of filaments \citep[e.g.,][]{lin04,lin05,lin07,lin08} clearly indicate that they are made by a myriad of thin ($0''.2 - 0''.6$) and long ($5'' - 20''$) thread-like structures piled up to form the filament body. These fine-structures, usually called fibrils or threads, seem to be orientated along magnetic field lines, and are observed in both active region and quiescent prominences. In observations from the Hinode spacecraft, threads usually appear as vertical structures in quiescent prominences \citep[e.g.,][]{berger}, while horizontal threads are more commonly seen in active region prominences \citep[e.g.,][]{okamoto}. However, this distinction does not seem very robust since the orientation of threads can significantly vary within the same filament \citep{lin04}. Theoretical models of filament threads \citep[e.g.,][]{ballesterpriest,rempel} usually represent them as part of larger magnetic flux tubes whose footpoints are anchored in the solar photosphere. The large number of observations of oscillations and propagating waves in filament threads \citep[see recent reviews by][]{oliverballester, ballester, banerjee, engvold} suggests that these phenomena are very frequent and ubiquitous in prominences. On the other hand, these oscillations are characterized by their quick damping, the damping time being of the order of a few, typically four, periods  \citep[the reader is referred to][and references therein, for a review of this issue]{oliver}.

Motivated by the observational evidence, a number of works studying filament thread oscillations from a theoretical point of view have been performed within the last decade. By considering the magnetohydrodynamic (MHD) theory, early works studied the MHD eigenmodes supported by a filament thread modeled as a Cartesian slab, partially filled with prominence plasma, and embedded in the corona \citep{joarder,diaz01,diaz03}. Later on, these works were extended by considering a more representative cylindrical geometry \citep{diaz02,dymovaruderman}. It is worth mentioning that the collective oscillations of systems of threads have been also investigated in both Cartesian \citep{diaz05,diazroberts} and cylindrical \citep{solercylinders} geometries. Focusing on the damping of the oscillations, studies on the effect of nonadiabatic mechanisms and mass flows \citep{solernonad} and resonant absorption \citep{arregui,solerslow} have been performed. All these referred works neglected the presence of neutrals and assumed a fully ionized hydrogen prominence plasma. Therefore, the effect of neutrals, and in particular that of ion-neutral collisions, on the propagation and damping of MHD waves in filament threads is not assessed yet and is the main motivation for the present investigation.

There is an extensive literature regarding wave propagation in a partially ionized multifluid plasma in the context of laboratory plasma physics \citep[e.g.,][]{watanabeA, watanabe,tanenbaum61,tanenbaum,woods,kulsrud,watts}. In astrophysical plasmas, the typical frequency of MHD waves is much smaller than the collisional frequencies between species. In such a case the one-fluid approach is usually adopted. One can find examples of works studying MHD waves in an unbounded, partially ionized, one-fluid plasma applied, e.g., to molecular clouds \citep{balsara}, to protoplanetary disks \citep{desch}, and to wave damping in the solar atmosphere \citep{depontieu,khoda04,leake05}. In the context of solar prominences, works by \citet{forteza,fortezanonad} are relevant. \citet{forteza} derived the full set of MHD equations for a partially ionized, one-fluid plasma and applied them to study the time damping of linear, adiabatic waves in an unlimited prominence medium. Subsequently, \citet{fortezanonad} extended their previous investigation to the nonadiabatic case by including thermal conduction by neutrals and electrons and radiative losses. Because of the effect of neutrals, in particular that of ion-neutral collisions, a generalized Ohm's law has to be considered, which causes some additional terms to appear in the resistive magnetic induction equation in comparison to the fully ionized case. Among these additional terms, the dominant one in the linear regime is the so-called ambipolar magnetic diffusion, which enhances magnetic diffusion across magnetic field lines. In this Paper, we apply the equations derived by \citet{forteza} to investigate the propagation of MHD waves in a partially ionized filament thread. Our filament thread model is composed of a homogeneous and infinite magnetic flux tube with prominence conditions surrounded by an unbounded and homogeneous coronal medium. The only previous investigation of waves affected by ion-neutral collisions in a solar magnetic structure is by \citet{kumarroberts}, but these authors considered the slab geometry and focused on surface waves in photospheric-like conditions. Hence, to our knowledge the present work is the first attempt to study MHD wave propagation in a partially ionized cylindrical flux tube.

This paper is organized as follows: Section~\ref{sec:math} contains a description of the model configuration and the basic equations. The results are presented and discussed in Section~\ref{sec:results}. Finally, our conclusions are given in Section~\ref{sec:conclusion}.

\section{MODEL AND METHOD}
\label{sec:math}

\subsection{Equilibrium Properties}

The model configuration considered in the present work is made of a homogeneous, isothermal, and infinite plasma cylinder of radius $a$, representing a filament thread, embedded in a unbounded coronal medium. We use cylindrical coordinates, namely $r$, $\varphi$, and $z$ for the radial, azimuthal, and longitudinal coordinates. In all the following expressions, a subscript  0 indicates local values, while subscripts $f$ and $c$ explicitly denote filament (internal) and coronal (external) quantities, respectively. The magnetic field is taken homogeneous and orientated along the $z$-direction, ${\mathbf B}_0 = B_0 \hat{\mathbf z}$, with $B_0 = 5$~G everywhere. By adopting the one-fluid approximation and considering only a hydrogen plasma \citep[see details in][]{forteza}, both the internal and external media are characterized by their plasma density, $\rho_0$, temperature, $T_0$, and number densities of neutrals, $n_{n}$, ions, $n_{i}$, and electrons $n_{e}$, with $n_{e}=n_{i}$. Thus, the gas pressure is $p_0 = (2 n_{i} + n_{n}) k_{\rm B} T_0$, where $k_{\rm B}$ is Boltzmann's constant. The relative densities of neutrals, $\xi_{n}$, and ions, $\xi_{i}$, are given by
\begin{equation}
 \xi_{n} = \frac{n_{n}}{n_{i} + n_{n}}, \qquad  \xi_{i} = \frac{n_{i}}{n_{i} + n_{n}},
\end{equation}
where we have neglected the contribution of electrons. We can now define a ionization fraction which gives us information about the plasma degree of ionization,
\begin{equation}
 \tilde{\mu} = \frac{1}{1+\xi_{i}}.
\end{equation}
This parameter is $\tilde{\mu} = 0.5$ for a fully ionized plasma and $\tilde{\mu} = 1$ for a neutral plasma. Any value of $\tilde{\mu}$ outside this range is physically meaningless. In the present work we take $T_f = 8000$~K, $\rho_f = 5 \times 10^{-11}$~kg~m$^{-3}$, and $\rho_c = \rho_f / 200 = 2.5 \times 10^{-13}$~kg~m$^{-3}$. The coronal medium is assumed to be fully ionized, so $ \tilde{\mu}_c = 0.5$. The filament ionization fraction, $ \tilde{\mu}_f$, is considered to be a free parameter, while the continuity of gas pressure selects the coronal temperature,
\begin{equation}
 T_c = \frac{\rho_f}{\rho_c} \frac{\tilde{\mu}_c}{ \tilde{\mu}_f} T_f. 
\end{equation}
Thus, the coronal temperature varies between $T_c = 8 \times 10^5$~K for  $ \tilde{\mu}_f = 0.5$ and $T_c = 1.6 \times 10^6$~K for $ \tilde{\mu}_f = 1$. MKS units are used throughout the paper.

\subsection{Basic Equations}

The detailed derivation of the basic MHD equations for a partially ionized hydrogen plasma in the one-fluid approach can be found in \citet{forteza}. After assuming small-amplitude, adiabatic perturbations from the equilibrium state, the basic equations are written in their linearized form \citep[see Equations~(18)--(21) of][]{forteza}, which in the notation adopted in the present work are
\begin{equation}
  \frac{\pd \rho_1}{\pd t} + \rho_0 \nabla \cdot {\mathit {\bf v}}_1 = 0, \label{eq:masscont}
\end{equation}
\begin{equation}
\rho_0  \frac{\pd {\mathit {\bf v}}_1}{\pd t}  = - \nabla p_1 + \frac{1}{\mu_0} \left[ \left( \nabla \times {\mathit {\bf B}}_1  \right) \times {\mathit {\bf B}}_0 \right], \label{eq:momentum}
\end{equation}
\begin{equation}
 \frac{\pd p_1}{\pd t} - \cs^2 \frac{\pd \rho_1}{\pd t} = 0,
\end{equation}
\begin{equation}
  \frac{\pd {\mathit {\bf B}}_1}{\pd t} = \nabla \times \left( {\mathit {\bf v}}_1 \times {\mathit {\bf B}}_0\right) + \eta \nabla^2 {\mathit {\bf B}}_1 + \eta_{\rm A} \nabla \times \left\{ \left[ \left( \nabla \times {\mathit {\bf B}}_1  \right) \times {\mathit {\bf B}}_0 - \frac{\mu_0 \xi_i}{1+\xi_i} \nabla p_1 \right] \times {\mathit {\bf B}}_0  \right\}, \label{eq:induction0}
\end{equation}
along with the condition $\nabla \cdot {\mathit {\bf B}}_1 = 0$. In these equations ${\mathit {\bf B}}_1 = \left( B_r , B_\varphi, B_z \right)$, ${\mathit {\bf v}}_1 = \left( v_r , v_\varphi, v_z \right)$, $\rho_1$, and $p_1$ are the magnetic field, velocity, density, and gas pressure perturbations, respectively, while $\cs^2 = \gamma p_0 / \rho_0$ and $\mu_0 = 4 \pi \times 10^{-7}$~N~A$^{-2}$ are the sound speed squared and the vacuum magnetic permeability, with $\gamma = 5/3$ the adiabatic index. Equations~(\ref{eq:masscont})--(\ref{eq:induction0}) are formally identical to their equivalent linear ideal MHD equations, with the exception of the induction equation (Equation~(\ref{eq:induction0})) which contains additional terms due to the presence of neutrals and a nonzero resistivity. Parameters $\eta$ and $\eta_{\rm A}$ are the coefficients of ohmic (also called Coulomb's) and ambipolar magnetic diffusion, respectively. In the present investigation, we neglect the Hall term in the induction equation because in a fully ionized plasma it may be important for frequencies larger than $\sim 10^4$ Hz, which are much larger than the observed frequencies of prominence oscillations. In a partially ionized plasma, the relative importance of the Hall effect grows with the density of neutrals, but in prominence conditions it can be still safely neglected \citep[see][]{pandey, krishan}. 

Equation~(\ref{eq:induction0}) can be rewritten in a more convenient form to explain the role of the nonideal terms. For this purpose, we define the quantity $\Xi$ and express the ambipolar diffusivity as,
\begin{equation}
 \Xi = \frac{\xi_n^2 \xi_i}{\left( 1 + \xi_i \right) \alpha_n}, \qquad  \eta_{\rm A} = \frac{\xi_n^2}{\mu_0 \alpha_n} =  \frac{\eta_{\rm C} - \eta}{\left| {\mathit {\bf B}}_0 \right|^2},
\end{equation}
where $\alpha_n$ is a friction coefficient and $\eta_{\rm C}$ is the so-called Cowling's coefficient of magnetic diffusion. Expressions for $\eta$, $\eta_{\rm C}$, and $\alpha_n$ as functions of the equilibrium physical properties and the ionization fraction are found in the literature \citep[e.g.,][]{bragi,khoda04,leake05,leake06} and are summarized in Appendix~\ref{app:parameters}. Thus, Equation~(\ref{eq:induction0}) becomes
\begin{equation}
 \frac{\pd {\mathit {\bf B}}_1}{\pd t} = \nabla \times \left( {\mathit {\bf v}}_1 \times {\mathit {\bf B}}_0\right) - \frac{1}{\sigma} \nabla \times {\mathit {\bf j}}_\parallel - \frac{1}{\sigma_ {\rm C}}  \nabla \times {\mathit {\bf j}}_\perp - \Xi \nabla \times \left( \nabla p_1 \times  {\mathit {\bf B}}_0 \right), \label{eq:induction}
\end{equation}
 where $\sigma =\left( \mu_0\eta \right)^{-1}$ and $\sigma_ {\rm C} = \left( \mu_0 \eta_{\rm C} \right)^{-1}$ are the ohmic and Cowling conductivities, whereas  ${\mathit {\bf j}}_\parallel$ and ${\mathit {\bf j}}_\perp$ are the parallel and perpendicular components  to the background magnetic field of the density current perturbation, ${\mathit {\bf j}}$, respectively, which are given by \citep{arber1}
\begin{equation}
 {\mathit {\bf j}}_\parallel = \frac{\left( {\mathit {\bf j}} \cdot {\mathit {\bf B}}_0\right){\mathit {\bf B}}_0}{\left|{ \mathit {\bf B}}_0 \right|^2}, \qquad
 {\mathit {\bf j}}_\perp = \frac{ {\mathit {\bf B}}_0 \times \left( {\mathit {\bf j}}\times {\mathit {\bf B}}_0 \right)}{\left|{ \mathit {\bf B}}_0 \right|^2},
\end{equation}
with $ {\mathit {\bf j}} = \left( \nabla \times  {\mathit {\bf B}}_1 \right)/\mu_0$. Hence, we see that parameters $\eta$ and $\eta_{\rm C}$ correspond to the coefficients of magnetic diffusion parallel and perpendicular to magnetic field lines, while the term with the factor $\Xi$ is responsible for coupling the gas pressure gradient to the magnetic field evolution. For a fully ionized plasma, $\eta_{\rm C} = \eta$ and $\Xi = 0$, so that ambipolar diffusion is suppressed and the magnetic diffusion is isotropic. In such a case, Equation~(\ref{eq:induction}) becomes
\begin{equation}
  \frac{\pd {\mathit {\bf B}}_1}{\pd t} = \nabla \times \left( {\mathit {\bf v}}_1 \times {\mathit {\bf B}}_0\right) - \eta \nabla \times \nabla \times{\mathit {\bf B}_1}, \label{eq:induction2}
\end{equation}
which corresponds to the usual induction equation for a fully ionized, resistive plasma. However, in the presence of neutrals $\eta_{\rm C} \neq \eta$, meaning that the main effect of ion-neutral collisions is to cause magnetic diffusion to become a nonisotropic process. It is worth mentioning that $\eta_{\rm C} \gg \eta$ even for a small relative density of neutrals. 

Since $\varphi$ and $z$ are ignorable coordinates, perturbed quantities are written proportional to $\exp \left( i  \omega t + i m \varphi - i k_z z \right)$, where $\omega$ is the frequency, and $m$ and $k_z$ are the azimuthal and longitudinal wavenumbers, respectively. After applying this Fourier-analysis to Equations~(\ref{eq:masscont})--(\ref{eq:induction0}), one obtains the following system of eight equations which govern the propagation of linear MHD waves,
\begin{equation}
 i \omega \rho_1 = -\rho_0 \left( v_r' + \frac{v_r}{r} + \frac{i m}{r} v_\varphi - i k_z v_z \right),  \label{eq:fouini}
\end{equation}
\begin{equation}
 i \omega  v_r = -\frac{1}{\rho_0}p_1' - \frac{\va^2}{B_0} \left( i k_z B_r + B_z' \right),
\end{equation}
\begin{equation}
  i \omega v_\varphi = - \frac{i m}{r} \frac{1}{\rho_0} p_1 - \frac{\va^2}{B_0} \left( i k_z B_\varphi + \frac{i m}{r} B_z  \right), \label{eq:alf1}
\end{equation}
\begin{equation}
  i \omega v_z = \frac{i k_z}{\rho_0} p_1,
\end{equation}
\begin{equation}
   i \omega  \left( p_1 - \cs^2 \rho_1 \right) = 0,
\end{equation}
\begin{equation}
  i \omega B_r =-i k_z B_0 v_r - \eta \left( \frac{m^2}{r^2} B_r + \frac{i m}{r} B_\varphi' +  \frac{i m}{r^2} B_\varphi  \right) - \eta_{\rm C} \left( k_z^2 B_r - i k_z B_z' \right) + i k_z \Xi B_0 p_1', \label{eq:ind1}
\end{equation}
\begin{equation}
 i \omega B_\varphi = - i k_z B_0 v_\varphi + \eta \left( B_\varphi'' + \frac{1}{r} B_\varphi' - \frac{1}{r^2} B_\varphi - \frac{i m}{r} B_r' + \frac{i m}{r^2} B_r\right) -  \eta_{\rm C} \left( k_z \frac{m}{r} B_z + k_z^2 B_\varphi \right) - k_z \Xi B_0 \frac{m}{r} p_1,  \label{eq:alf2}
\end{equation} 
\begin{eqnarray}
  i \omega B_z &=&- B_0\left( v_r' + \frac{1}{r} v_r + \frac{i m}{r} v_\varphi \right) +  \eta_{\rm C} \left( B_z'' + \frac{1}{r} B_z' - \frac{m^2}{r^2} B_z + i k_z B_r' + \frac{i k_z}{r} B_r - k_z \frac{m}{r} B_\varphi \right) \nonumber \\ & &+ \Xi B_0 \left( p_1'' + \frac{1}{r} p_1' -  \frac{m^2}{r^2} p_1  \right),\label{eq:foufin}
\end{eqnarray}
where the prime denotes derivative with respect to $r$ and $\va^2 = B_0^2 / \mu_0 \rho_0$ is the Alfv\'en speed squared. Equations~(\ref{eq:fouini})--(\ref{eq:foufin}) form an eigenvalue problem which we numerically solve by means of the PDE2D code \citep{sewell} based on finite elements \citep[see, for example,][for an explanation of the method]{terradas}. The numerical integration of Equations~(\ref{eq:fouini})--(\ref{eq:foufin}) is performed from the cylinder axis, $r=0$, to the finite edge of the numerical domain, $r=r_{\rm max}$. The evanescent condition is imposed in the coronal medium, so all perturbations vanish at $r=r_{\rm max}$. We impose the evanescent condition because we focus our study on trapped, nonleaky modes supported by the filament thread body. It is unlikely that the observed thread oscillations correspond to leaky modes, which are high-frequency propagating waves in the coronal medium  \citep[see for example][]{cally}, while an interpretation in terms of trapped solutions is more realistic and consistent with the observed periods. Therefore, the edge of the numerical domain has been located far enough from the filament thread to obtain a good convergence of the solution and to avoid numerical errors (typically, we consider $r_{\rm max} > 100 a$). The boundary conditions at $r=0$ are imposed by symmetry arguments. We assume a fixed, real, and positive $k_z$, so the numerical solution provides with a complex oscillatory frequency,  $\omega$, as well as the eigenfunctions of perturbations. Wave solutions appear in pairs, $\omega_\pm = \pm \omega_{\rm R} + i \omega_{\rm I}$. The solution $\omega_+$ corresponds to a wave propagating towards the positive $z$-direction, whereas the contrary stands for $\omega_-$. Since there are no flows in the equilibrium configuration, both solutions have the same properties, and so we restrict ourselves to solutions with $\Re \left( \omega \right) > 0$. The oscillatory period, $P$, the damping time, $\tau_{\rm D}$, and the damping ratio, $\tau_{\rm D} /P$, are related to the frequency as follows
\begin{equation}
 P = \frac{2 \pi}{\omega_{\rm R}}, \quad \tau_{\rm D} = \frac{1}{\omega_{\rm I}}, \quad \frac{\tau_{\rm D}}{P} =\frac{1}{2\pi} \frac{\omega_{\rm R}}{\omega_{\rm I}}.
\end{equation}

\subsection{Dimensional Analysis of the Induction Equation}

\citet[][see their Figure~ 6]{forteza} evaluated the relative importance of the nonideal terms of the induction equation for typical prominence conditions. For the range of parameters considered in their paper, these authors concluded that the effect of the pressure-related term, i.e., that with the parameter $\Xi$, is negligible, while the term with Cowling's diffusivity is the dominant one. Here we quantify the importance of ohmic and Cowling's diffusion by means of a dimensional analysis. Equation~(\ref{eq:induction}) indicates that magnetic diffusion in the parallel and perpendicular directions is governed by the terms with ohmic diffusivity and Cowling's diffusivity, respectively. We thus define the parallel and perpendicular magnetic Reynolds numbers as follows,
\begin{equation}
 R_{m \parallel} = \frac{U_0 L_\eta^2}{\eta L}, \qquad  R_{m \perp} = \frac{U_0 L_{\eta_{\rm C}}^2 }{\eta_{\rm C} L}, \label{eq:reynolds}
\end{equation}
where $L_\eta$ and $L_{\eta_{\rm C}}$ are typical length-scales related to the terms of the induction equation with $\eta$ and  $\eta_{\rm C}$, respectively, whereas $L = \min \left( L_\eta, L_{\eta_{\rm C}} \right)$, and $U_0$ is a typical plasma velocity. In the absence of neutrals, $L_\eta = L_{\eta_{\rm C}} = L$, so $R_{m \parallel} = R_{m \perp} =  \frac{U_0 L}{\eta}$ because magnetic diffusion is isotropic. However, since this mechanism is nonisotropic in a partially ionized plasma, both the longitudinal and perpendicular directions have different typical length-scales. An examination of Equations~(\ref{eq:ind1}), (\ref{eq:alf2}), and (\ref{eq:foufin}) reveals that longitudinal derivatives only appear in the terms with $\eta_{\rm C}$, whereas those with $\eta$ only contain radial and azimuthal derivatives. Hence, $L_\eta$ is a typical length-scale perpendicular to magnetic field lines and can be related to the filament thread radius, $L_\eta \sim a$. On the other hand, $L_{\eta_{\rm C}}$ corresponds to a typical length-scale along the magnetic field direction and can be expressed in terms of the longitudinal wavelength, $L_{\eta_{\rm C}} \sim \lambda_z$ or, equivalently, in terms of the longitudinal wavenumber, $L_{\eta_{\rm C}} \sim 2\pi k_z^{-1}$. In addition, we relate the velocity-scale to the filament thread sound speed, $U_0 \sim \csp$. Therefore, the Reynolds numbers (Equation~(\ref{eq:reynolds})) are rewritten as follows,
\begin{equation}
  R_{m \parallel} = \frac{\csp a}{\eta}, \qquad  R_{m \perp} = \frac{4 \pi^2 \csp }{\eta_{\rm C} k_z^2 a}.
\end{equation}
We see that $R_{m \parallel}$ is independent of the longitudinal wavenumber, while $R_{m \perp}$ is inversely proportional to $k_z^2$. This suggests that the relative importance of Cowling's diffusion increases with $k_z$. To perform a simple calculation we take $k_z a \sim 1$, consider a typical value $a = 100$~km, and assume $ \tilde{\mu}_f = 0.8$. We obtain $R_{m \parallel} \approx 7 \times 10^6$ and $R_{m \perp} \approx 4 \times 10^2$. For the considered parameters, the dominant nonideal mechanism is Cowling's diffusion, while ohmic diffusion has a minor role. Furthermore, it is possible to obtain an estimation of the wavenumber for which both ohmic and Cowling's diffusion have the same importance by equaling the parallel and perpendicular magnetic Reynolds numbers, i.e., $R_{m \parallel} = R_{m \perp}$. By this procedure we obtain,
\begin{equation}
 k_z a \approx 2 \pi \sqrt{\frac{\eta}{\eta_{\rm C}}}, \label{eq:kztransestimated}
\end{equation}
which gives $k_z a \approx 2.8 \times 10^{-2}$ for the same parameters assumed before. These estimations are consistent with \citet{forteza}, since they considered quite a large value for the longitudinal wavenumber and therefore ohmic diffusion is not relevant in their case. We have to bear in mind that the observed width of filament threads is in the range $0''.2 - 0''.6$ \citep{lin04}, and therefore $a$ ranges from 75~km to 375~km, approximately. On the other hand, the detected wavelengths of prominence oscillations are between $5 \times 10^3$~km and $10^5$~km \citep{oliverballester}. One can combine both quantities (the wavelength and the thread width) into the dimensionless quantity $k_z a$ and compute its upper and lower limits. So, for $a=75$~km, $5 \times 10^{-3} < k_z a < 9 \times 10^{-2}$, while for $a=375$~km, $2 \times 10^{-2} < k_z a < 4 \times 10^{-1}$. Thus, taking into account both intervals and considering that thinner threads than those resolved by present-day telescopes might exist, the relevant range of $k_z a$ of prominence oscillations covers two orders of magnitude and corresponds to $10^{-3} < k_z a < 10^{-1}$. This range of $k_z a$ contains all realistic values of the wavelength and the thread width. According to Equation~(\ref{eq:kztransestimated}), both ohmic and Cowling's diffusion could be important in such a range of $k_z a$. This will be verified through the numerical computations in the next Sections.

\section{RESULTS}
\label{sec:results}

\subsection{Alfv\'en Waves}
\label{sec:alfven}

\begin{figure*}[!htp]
\centering
\epsscale{1}
\plotone{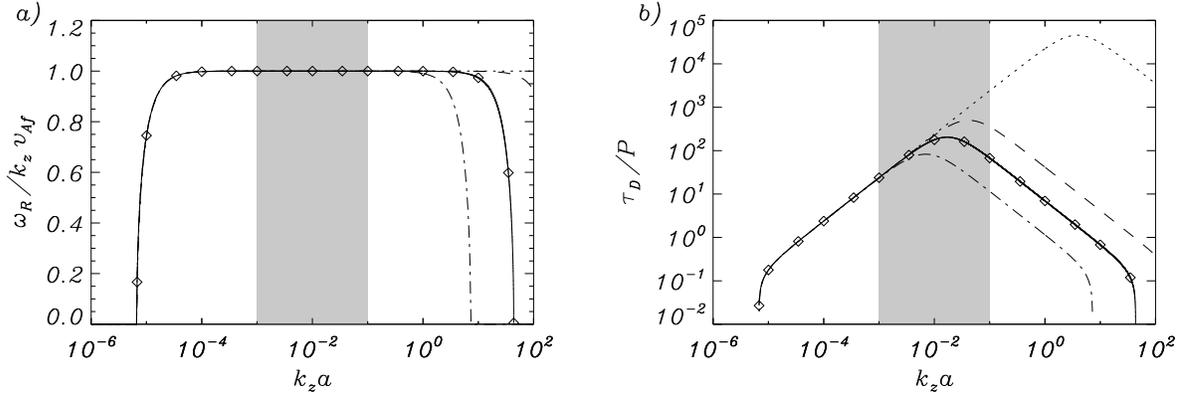}
\caption{Results corresponding to the Alfv\'en mode. ($a$) Phase speed as a function of $k_z a$ in units of the internal Alfv\'en speed. ($b$) Ratio of the damping time to the period as a function of $k_z a$. In both panels, different linestyles represent different ionization degrees: $\tilde{\mu}_f = 0.5$ (dotted line),  $\tilde{\mu}_f = 0.6$ (dashed line), $\tilde{\mu}_f = 0.8$ (solid line), and $\tilde{\mu}_f = 0.95$ (dash-dotted line). Symbols are the approximate solution given by Equation~(\ref{eq:alfventot}) for $\tilde{\mu}_f = 0.8$. The shaded zone corresponds to the range of typically observed wavelengths of prominence oscillations. \label{fig:alfven}}
\end{figure*}

\begin{figure}[!ht]
\centering
\epsscale{0.5}
\plotone{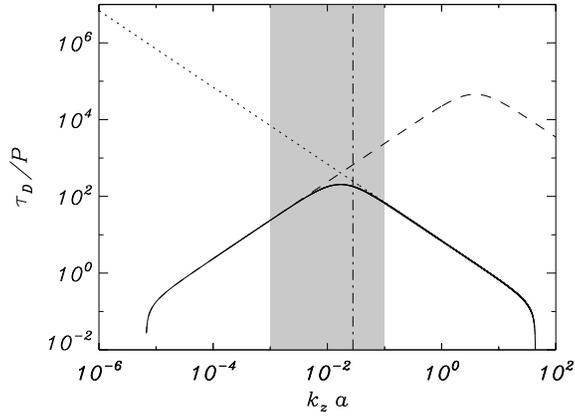}
\caption{Ratio of the damping time to the period as a function of $k_z a$ corresponding to the Alfv\'en mode with $\tilde{\mu}_f = 0.8$. The solid line is the complete solution considering all terms in the induction equation, i.e., the solid line in Figure~\ref{fig:alfven}($b$). The dotted and dashed lines are the results obtained by neglecting ohmic diffusion ($\eta = 0$) or ion-neutral collisions ($\eta_{\rm C} = \eta$), respectively. The vertical dot-dashed line is the analytically estimated transitional wavenumber between both regimes of dominance (Equation~(\ref{eq:kztransestimated})). \label{fig:alfvendamping}}
\end{figure}

We start our investigation by studying Alfv\'en waves. Some works have investigated Alfv\'en wave propagation in partially ionized plasmas \citep[e.g.,][]{watanabe,tanenbaum,watts,pandey, fortezanonad}, but to our knowledge a detailed investigation in cylindrical geometry applied to solar plasmas remains to be done. In general Alfv\'en modes are coupled to magnetoacoustic modes except for $m=0$. Thus, here we assume no azimuthal dependence in order to study Alfv\'en waves separately. We see that by setting $m=0$ Equations~(\ref{eq:alf1}) and (\ref{eq:alf2}) are decoupled from the rest,
\begin{equation}
  i \omega v_\varphi =- i k_z \frac{\va^2}{B_0} B_\varphi, \label{eq:alfven1m0}
\end{equation}
\begin{equation}
  i \omega B_\varphi = - i k_z B_0 v_\varphi + \eta \left( B_\varphi'' + \frac{1}{r} B_\varphi' -\frac{1}{r^2} B_\varphi \right) -  \eta_{\rm C}  k_z^2 B_\varphi.  \label{eq:alfven2m0}
\end{equation}
These equations only involve the perturbations $v_\varphi$ and $B_\varphi$, so Alfv\'en modes are purely torsional, incompressible waves for $m=0$. Now, we use Equation~(\ref{eq:alfven1m0}) to express $v_\varphi$ as a function of  $B_\varphi$, and after substituting it in Equation~(\ref{eq:alfven2m0}) a single equation for $B_\varphi$ is obtained,
\begin{equation}
 \eta B_\varphi'' + \eta \frac{1}{r} B_\varphi' + \left[ \frac{i}{\omega} \left(  k_z^2 \va^2 - \omega^2 \right) - \eta_{\rm C} k_z^2 - \eta \frac{1}{r^2} \right] B_\varphi = 0. \label{eq:alfvenfin1}
\end{equation}
Equation~(\ref{eq:alfvenfin1}) can be rewritten as follows,
\begin{equation}
r^2  B_\varphi'' + r B_\varphi' + \left( m_{\rm A}^2 r^2 - 1 \right) B_\varphi = 0, \label{eq:alfvenbessel}
\end{equation}
with
\begin{equation}
 m_{\rm A}^2 = \frac{i}{\eta \omega} \left( k_z^2 \Gamma_{\rm A}^2 - \omega^2\right), \qquad  \Gamma_{\rm A}^2 = \va^2 + i \omega\eta_{\rm C}. \label{eq:gamma}
\end{equation}
The same equation stands for $v_\varphi$. So, the Alfv\'en wave is governed by a Bessel equation of order 1, where $m_{\rm A}$ plays the role of the radial wavenumber and $\Gamma_{\rm A}$ is equivalent to the modified Alfv\'en speed defined in Equation~(29) of \citet{fortezanonad}. We see then that in the resistive case, Alfv\'en modes are not strictly confined to magnetic surfaces and become ``global'' eigenmodes of the flux tube \citep[see][]{ferraro,sy,copil}. The general solution of Equation~(\ref{eq:alfvenbessel}) for regular perturbations at $r=0$ and vanishing at infinity is
\begin{equation}
 B_\varphi (r) = \left\{ \begin{array}{lcl} 
                 A_1 J_1 (m_{{\rm A}f} r ) & {\rm if} & r \le a, \\
		 A_2 H_1^{(1)} (m_{{\rm A}c} r ) & {\rm if} & r > a,
                \end{array} \right. \label{eq:bphiprofile}
\end{equation}
$A_1$ and $A_2$ being complex constants. $J_1$ and $H_1^{(1)}$ are the usual Bessel function and Hankel function of the first kind, respectively, of order 1 \citep{abramowitz}. In order to obtain the dispersion relation, we need to impose boundary conditions at $r=a$. \citet{tonimura} \citep[see also][]{woods} provide appropriate boundary conditions for our case,
\begin{equation}
 \left[ \left[ B_\varphi \right]\right] = 0, \qquad  \left[ \left[ \eta B_\varphi' \right]\right] = - \frac{B_\varphi}{r}  \left[ \left[ \eta \right] \right],
\end{equation}
where $\left[ \left[ X \right]\right] = X_c - X_f$ stands for the jump of the quantity $X$ at $r=a$. $B_\varphi$ is set continuous across the boundary because we assume no surface currents on the interface between the thread and the corona. On the contrary, the jump of the derivative of  $B_\varphi$ is determined by the difference between the internal and external ohmic diffusivities. Applying these boundary conditions and after some algebra, the following dispersion relation is obtained,
\begin{equation}
 \eta_c  m_{{\rm A}c}  \frac{{H'_1}^{(1)} (m_{{\rm A}c} a ) }{H_1^{(1)} (m_{{\rm A}c} a ) } - \eta_f m_{{\rm A}f} \frac{J_1' (m_{{\rm A}f} a ) }{J_1 (m_{{\rm A}f} a) } = \frac{\eta_f  - \eta_c}{a}. \label{eq:disp1}
\end{equation}
Next, we solve Equation~(\ref{eq:disp1}) and plot in Figure~\ref{fig:alfven}($a$) the Alfv\'en wave phase speed as a function of $k_z a$ for several ionization degrees. As one can see, the Alfv\'en mode only propagates between two critical wavenumbers, the first of them being independent from the ionization degree. At these critical wavenumbers, the real part of the Alfv\'en frequency vanishes. Note that these critical values are far from the range of $k_z a$ relevant for prominence oscillations. On the other hand, Figure~\ref{fig:alfven}($b$) displays the ratio $\tau_{\rm D}/P$ as a function of $k_z a$. $\tau_{\rm D}/P$ is independent from $\tilde{\mu}_f$ for small $k_z a$, while it is significantly affected by the ionization degree for large $k_z a$. The maximum of $\tau_{\rm D}/P$ occurs within or close to the relevant range of $k_z a$. Figure~\ref{fig:alfvendamping} allows a better understanding of the behavior of $\tau_{\rm D}/P$. Here, we display a comparison of the actual value of $\tau_{\rm D}/P$ with that obtained by neglecting one of the two possible damping mechanisms, i.e., ohmic diffusion (by setting $\eta = 0$) or ion-neutral collisions (by setting $\eta_{\rm C} = \eta$). As expected, the presence of neutrals has an important effect on the damping time for large $k_z a$, whereas ohmic diffusion dominates for small $k_z a$. As estimated by Equation~(\ref{eq:kztransestimated}), the transition between both regimes takes place within the relevant range of $k_z a$, and the estimated transitional wavenumber is close to the actual value.

To perform a more in-depth study of Alfv\'en wave propagation, let us consider now the case in which ohmic diffusion can be neglected, i.e., $k_z a$ much larger than the transitional one (Equation~(\ref{eq:kztransestimated})). So, by taking $\eta = 0$ in Equation~(\ref{eq:alfvenfin1}) one obtains that Alfv\'en waves verify,
\begin{equation}
 \left[ \frac{i}{\omega} \left(k_z^2 \va^2 - \omega^2 \right) - \eta_{\rm C} k_z^2 \right] B_\varphi = 0,
\end{equation}
which for an arbitrary $B_\varphi$ gives us the dispersion relation for Alfv\'en modes in the regime dominated by Cowling's diffusion,
\begin{equation}
 \omega^2 - i \eta_{\rm C} k_z^2 \omega - k_z^2 \va^2 = 0. \label{eq:alfvendisper1}
\end{equation}
Equation~(\ref{eq:alfvendisper1}) is formally identical to Equation~(49) of \citet{pandey}. Its exact solution gives us a complex frequency,
\begin{equation}
 \omega = \pm \frac{k_z}{2} \sqrt{4 \va^2 - \eta_{\rm C}^2 k_z^2} + i \frac{k_z^2}{2} \eta_{\rm C}. \label{eq:alfvenfreqcouling}
\end{equation}
Because of the presence of an imaginary part of the frequency, we obtain the well-known result that the Alfv\'en wave is damped in a partially ionized plasma \citep[see, e.g.,][for studies on the damping of Alfv\'en waves in the solar chromosphere]{haerendel92,depontieuhae,depontieu99,depontieu, leake05}. Obviously, if $\eta_{\rm C} = 0$ we recover the ideal, undamped Alfv\'en mode, $\omega =  \pm k_z \va$. From Equation~(\ref{eq:alfvenfreqcouling}) we can also see that the real part of the Alfv\'en frequency vanishes for a critical value of $k_z$,
\begin{equation}
 k_{z}^{c} = \frac{2 \va}{\eta_{\rm C}}. \label{eq:crit1}
\end{equation}
This critical wavenumber is equivalent to that given by Equation~(38) of \citet{fortezanonad} by considering parallel propagation. For $k_z > k_{z}^c$ the Alfv\'en wave becomes a nonpropagating, purely damped disturbance. This critical wavenumber corresponds to the largest critical wavenumber obtained in Figure~\ref{fig:alfven}. 

\begin{figure}[!ht]
\centering
\epsscale{0.5}
\plotone{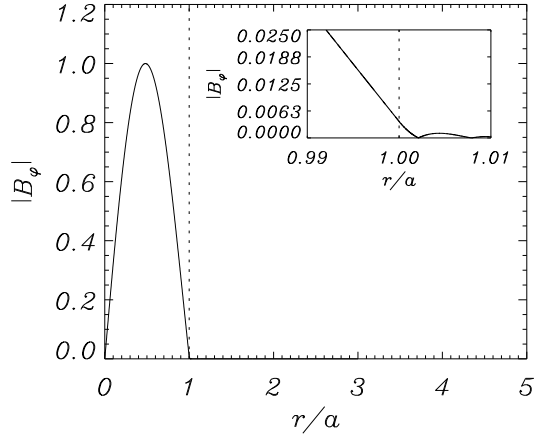}
\caption{Modulus in arbitrary units of the eigenfunction $B_{\varphi}$ as a function of $r/a$ corresponding to the radially fundamental Alfv\'en mode with $m=0$, $\tilde{\mu}_f = 0.8$, and $k_z a = 10^{-2}$. The vertical dotted line corresponds to the filament thread edge. The small panel shows a close-up of the eigenfunction close to the boundary between the thread and the corona. \label{fig:bphi}}
\end{figure}

To obtain an expression for the smaller critical wavenumber, we have to consider the case $\eta \neq 0$. To make some analytical progress we start by plotting the $B_\varphi$ perturbation corresponding to the radially fundamental Alfv\'en mode (Figure~\ref{fig:bphi}). Although the eigenfunction is not strictly confined within the cylinder, we note that the amplitude of $B_\varphi$ for $r > a$ is much smaller than that within the thread. This is caused by the large contrasts of the internal and external densities, $\rho_f / \rho_c = 200$, and ohmic diffusivities, $\eta_f / \eta_c \approx 3 \times 10^3$. On the basis of this evidence, it is possible to give an analytical approximation of the Alfv\'en frequency in the general $\eta \neq 0$ case by neglecting the influence of the corona and setting $B_\varphi \approx 0$ at $r=a$. This implies that $m_{{\rm A}f} a \approx j_1$, with $ j_1$ the first zero of the Bessel function $J_1$. In this approximation, the Alfv\'en mode frequency is 
\begin{equation}
 \omega = \pm \frac{1}{2} \sqrt{4 \va^2 k_z^2 - \left( \eta_{\rm C} k_z^2 + \eta \frac{j_1^2}{a^2} \right)^2} + \frac{i}{2} \left( \eta_{\rm C} k_z^2+ \eta \frac{j_1^2}{a^2} \right). \label{eq:alfventot}
\end{equation}
This expression is consistent with previous results since for $\eta = 0$ it reduces to Equation~(\ref{eq:alfvenfreqcouling}). As before, we can obtain an expression for the critical $k_z$ at which the real part of the Alfv\'en frequency vanishes,
\begin{equation}
  k_{z}^{c\pm} = \frac{\va}{\eta_{\rm C}} \pm \frac{\sqrt{\va^2 - \eta_{\rm C} \eta j_1^2/a^2}}{\eta_{\rm C}}. \label{eq:kzcritfull}
\end{equation}
We see that two $k_z^c$ are now possible. That given by the $+$ sign, namely $k_z^{c+}$, is a correction of the previously described critical wavenumber for $\eta = 0$ (Equation~(\ref{eq:crit1})). On the other hand, that given by the $-$ sign, namely $k_z^{c-}$, is a new critical value which arises because of the combination of two effects, namely the nonzero value of $\eta$ and the presence of the factor $j_1^2/a^2$ given by the geometry. Since \citet{fortezanonad} considered an infinite medium, this geometry-related critical wavenumber is absent in their investigation. Considering that $\eta_{\rm C} \eta j_1^2/a^2 \va^2 \ll 1$, a first-order Taylor expansion of Equation~(\ref{eq:kzcritfull}) gives
\begin{equation}
 k_z^{c+} \approx \frac{\va}{\eta_{\rm C}} \left( 2 -  \frac{\eta_{\rm C} \eta j_1^2/a^2}{2 \va^2}\right) \approx \frac{2\va}{\eta_{\rm C}} , \qquad  k_z^{c-} \approx \frac{\eta j_1^2/a^2}{2 \va}. \label{eq:alfvencriticsmasmenos}
\end{equation}
One can see that $k_z^{c-}$ does not depend on $\eta_{\rm C}$ and so it is not affected by the plasma ionization degree. Furthermore, since $k_z^{c-} < k_z^{c+}$, the Alfv\'en wave only exists as a propagating mode for $k_z^{c-} < k_z < k_z^{c+}$.

For $\tilde{\mu}_f = 0.8$ the approximation given by Equation~(\ref{eq:alfventot}) is compared with the complete solution by means of symbols in Figure~\ref{fig:alfven}. We see that there is a good agreement between both solutions meaning that both the approximation to the frequency (Equation~(\ref{eq:alfventot})) and the expressions for the critical wavenumbers (Equation~(\ref{eq:alfvencriticsmasmenos})) are consistent with the actual result. The Alfv\'en wave phase speed is identical to $\va$ in a fully ionized, ideal plasma. Therefore, we can perform an analogy and define an effective internal Alfv\'en speed from the real part of Equation~(\ref{eq:alfventot}),
\begin{equation}
 \tilde{v}_{{\rm A}f} = \sqrt{\vap^2 - \frac{k_z^2}{4} \left( \eta_{\rm C} + \eta \frac{j_1^2}{k_z^2 a^2} \right)^2}. \label{eq:effectiveva}
\end{equation}
We see that except when $k_z$ is close to one of the critical wavenumbers, the phase speed is almost equal to $\vap$.

\subsection{Kink Mode}

Here, we turn our attention to the kink mode. We fix the azimuthal wavenumber to $m=1$ since we are interested in modes that are able to perturb the cylinder axis and are reasonable candidates to produce the observed transverse oscillations of prominence threads \citep[e.g.,][]{okamoto}. We focus on the radially fundamental kink mode. It is not possible to give an analytical dispersion relation for this wave when both ohmic and ambipolar magnetic diffusion are considered in the induction equation. For this reason, we numerically solve the full eigenvalue problem (Equations~(\ref{eq:fouini})--(\ref{eq:foufin})) and obtain the frequency with the PDE2D code.

\begin{figure*}[!htp]
\centering
\epsscale{1}
\plotone{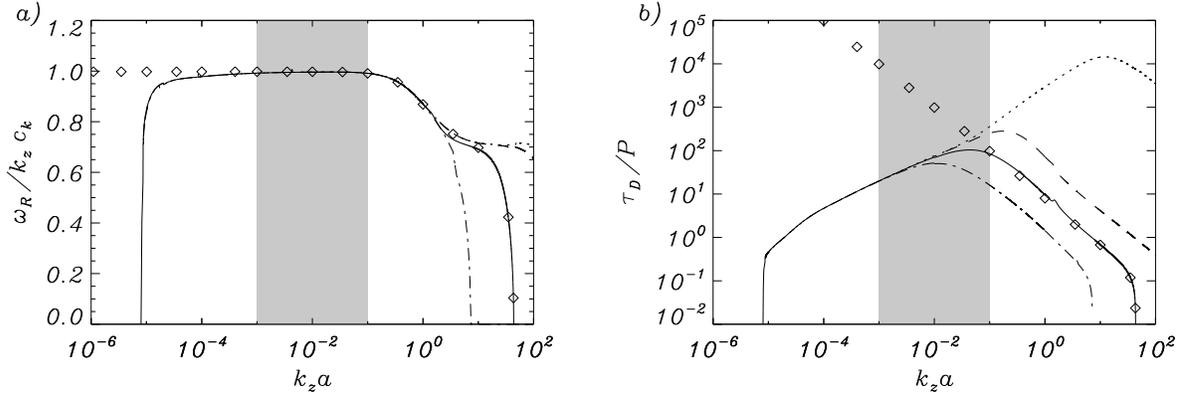}
\caption{Results corresponding to the kink mode. ($a$) Phase speed as a function of $k_z a$ in units of the kink speed. ($b$) Ratio of the damping time to the period as a function of $k_z a$. The different linestyles and the shaded zone have the same meaning as in Figure~\ref{fig:alfven}. Symbols are the approximate solution given by solving Equation~(\ref{eq:disperapproxkink}) for $\tilde{\mu}_f = 0.8$. \label{fig:kink}}
\end{figure*}

Figure~\ref{fig:kink}($a$) displays the kink mode phase speed as a function of $k_z a$. We see that the propagation of the kink wave is also constrained by the existence of two critical wavenumbers, that turn out to be the same found in the case of the Alfv\'en wave, see Section~\ref{sec:alfven}. This behavior can be easily understood by considering the first-order asymptotic approximations of the kink mode frequency in the ideal case \citep{edwinroberts},
\begin{equation}
  \omega \approx \left\{ \begin{array}{lcl} 
		 c_k k_z & {\rm for} & k_z a \ll 1, \\
		\vap k_z & {\rm for} & k_z a \gg 1, 
                \end{array} \right.
\end{equation}
where $c_k = \sqrt{\frac{2}{1+\rho_c / \rho_f }} \vap$ is the so-called kink speed. Note that for $\rho_f \gg \rho_c$, $c_k \approx \sqrt{2} \vap$. So, both in the short- and long-wavelength regimes the kink mode frequency directly depends on the Alfv\'en speed. Replacing the ideal Alfv\'en speed by its effective value, $\tilde{v}_{{\rm A}f}$, from Equation~(\ref{eq:effectiveva}), one sees that the kink mode frequency vanishes at the same critical wavenumbers as the Alfv\'en wave. Drawing our attention within the relevant range of $k_z a$, we obtain $\omega / k_z \approx c_k$, so nonideal effects do not seem relevant to wave propagation for the observed wavelengths. Next, the dependence of the ratio of the damping time to the period with $k_z a$ is shown in Figure~\ref{fig:kink}($b$). This result is very similar to that obtained for the Alfv\'en mode, hence no additional comments are needed in this case.

Some analytical progress can be performed by adopting the $\beta = 0$ case and neglecting ohmic diffusion. In such a situation, from Equation~(\ref{eq:momentum}) one can write the magnetic force term as,
\begin{equation}
\left( \nabla \times {\mathit {\bf B}}_1  \right) \times {\mathit {\bf B}}_0 = \frac{i \omega B_0^2}{\va^2} {\mathit {\bf v}}_1,
\end{equation}
which can be used to rewrite Equation~(\ref{eq:induction0}) in a compact form as follows,
\begin{equation}
  \frac{\pd {\mathit {\bf B}}_1}{\pd t} = \frac{\Gamma_{\rm A}^2}{\va^2} \nabla \times \left( {\mathit {\bf v}}_1 \times {\mathit {\bf B}}_0\right), \label{eq:inductionapprox}
\end{equation}
with $\Gamma_{\rm A}^2$ defined in Equation~(\ref{eq:gamma}). Equation~(\ref{eq:inductionapprox}) is formally identical to the ideal induction equation with the extra factor $\Gamma_{\rm A}^2 / \va^2$ in the convective term. Note that $\Gamma_{\rm A}^2 / \va^2 = 1$ for $\eta_{\rm C} = 0$. Thus, it is now straight-forward to obtain a dispersion relation by following the same procedure as in the ideal case \citep[see details in][]{edwinroberts},
\begin{equation}
\rho_c  m_f \left( \omega^2 - k_z^2 \Gamma_{{\rm A}c}^2 \right) \frac{J'_m \left( m_f a \right)}{J_m \left( m_f a \right)} 
 = \rho_f  m_c \left( \omega^2 - k_z^2 \Gamma_{{\rm A}f}^2 \right) \frac{K'_m \left( m_c a \right)}{K_m \left( m_c a \right)}, \label{eq:disperapproxkink}
\end{equation}
where $K_m$ is the modified Bessel function and the quantities $m_f$ and $m_c$ are given by
\begin{equation}
m_f^2 = \frac{\left( \omega^2 - k_z^2 \Gamma_{{\rm A}f}^2  \right)  }{\Gamma_{{\rm A}f}^2}, \qquad  m_c^2 = \frac{\left(k_z^2 \Gamma_{{\rm A}c}^2 - \omega^2  \right)  }{\Gamma_{{\rm A}c}^2}.
\end{equation}
Note that Equation~(\ref{eq:disperapproxkink}) applies to any value of $m$. Equation~(\ref{eq:disperapproxkink}) has been solved for $m=1$ and $\tilde{\mu}_f = 0.8$, the corresponding results being the symbols in Figure~\ref{fig:kink}. This approximate analytical solution agrees well with the numerical solution in the range of $k_z a$ dominated by Cowling's diffusion. For small $k_z a$, the numerical and analytical solutions do not agree because the effect of ohmic diffusion is missed by the analytical approximation. 

\begin{figure}[!ht]
\centering
\epsscale{0.5}
\plotone{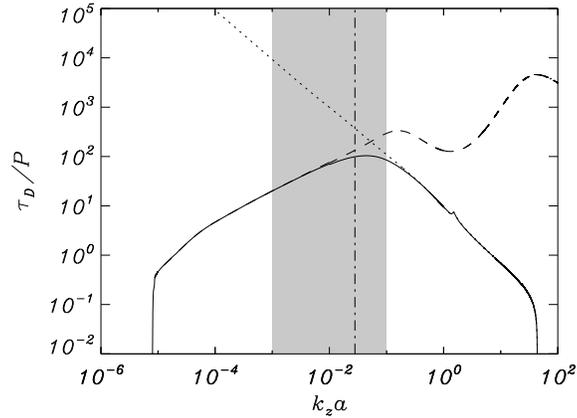}
\caption{Same as Figure~\ref{fig:alfvendamping} but for the kink mode. \label{fig:dampingkink}}
\end{figure}

Next, we check here the efficiency of the nonideal terms of the induction equation on the kink mode damping. Figure~\ref{fig:dampingkink} shows a comparison of the value of $\tau_{\rm D}/P$ of the kink wave with that obtained by neglecting ohmic diffusion ($\eta = 0$) or ion-neutral collisions ($\eta_{\rm C} = \eta$ and $\Xi = 0$). Only the solutions corresponding to $\tilde{\mu}_f = 0.8$ are displayed. The result for the kink mode is very similar to that obtained for the Alfv\'en mode (see Figure~\ref{fig:alfvendamping}): ohmic diffusion dominates for small $k_z a$ and ion-neutral collisions are dominant for large $k_z a$. The transition between both behaviors takes place within the relevant range of $k_z a$, where  $\tau_{\rm D}/P \gg 4$. Therefore, neither ohmic diffusion nor ion-neutral collisions can provide damping times compatible with those observed. Only for an almost neutral plasma ($\tilde{\mu}_f > 0.95$) and large $k_z$, one can obtain a $\tau_{\rm D}/P$ consistent with the observations. In comparison with the damping mechanisms studied in previous works, the efficiency of ohmic diffusion and ion-neutral collisions is much smaller than that of resonant absorption for the kink mode damping \citep{arregui,solerslow}.

\subsection{Slow Mode}

\begin{figure*}[!htp]
\centering
\epsscale{1}
\plotone{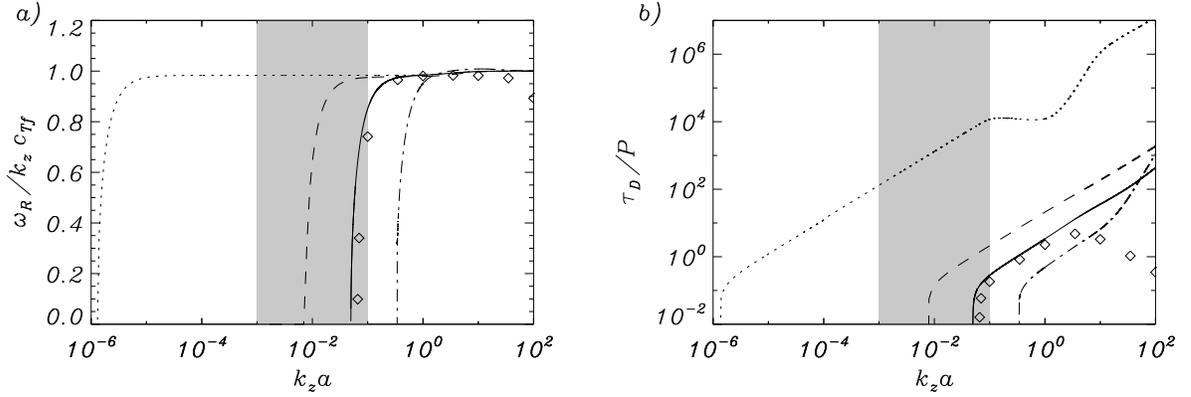}
\caption{Results corresponding to the slow mode with $m=1$. ($a$) Phase speed as a function of $k_z a$ in units of the internal cusp speed speed. ($b$) Ratio of the damping time to the period as a function of $k_z a$. The different linestyles and the shaded zone have the same meaning as in Figure~\ref{fig:alfven}. Symbols correspond to the analytical approximate solution for $\tilde{\mu}_f = 0.8$ (Equation~(\ref{eq:approxslow})). \label{fig:slow}}
\end{figure*}

The last wave mode that we study in this investigation is the slow mode. Again we numerically obtain the frequency of the radially fundamental mode. We consider $m=1$ but we note that the slow mode behavior is weakly affected by the value of the azimuthal wavenumber. Figure~\ref{fig:slow}($a$) shows the slow mode phase speed as a function of $k_z a$. The slow mode behavior is also affected by the presence of a critical wavenumber which is highly dependent on the ionization degree. The slow mode is totally damped for $k_z$ smaller than the critical value. We see that for large enough $\tilde{\mu}_f$, the critical $k_z$ falls inside or is larger than the observed values. This result has relevant implications from the observational point of view, since it suggests that the slow wave might not propagate in realistic, thin filament threads. For $k_z a$ larger than the critical value, the slow mode phase speed is close to the internal tube (or cusp) speed, $\ctp = \csp \vap / \sqrt{\csp^2 + \vap^2 }$. On the other hand, Figure~\ref{fig:slow}($b$) displays $\tau_{\rm D} / P$ again as a function of $k_z a$, where one can see that the damping time decreases dramatically as $k_z a$ approaches the critical value. Also, we see that the damping time grows rapidly when $\tilde{\mu}_f \to 0.5$, i.e., the dependence on the ionization degree is more sensitive to the ionization degree for an almost fully ionized plasma than for a weakly ionized plasma.

Such as for the Alfv\'en and kink waves, it is possible to perform some simple analytical calculations in order to get a better understanding of the slow mode behavior. Here, it is crucial to take into account the common result regarding the ideal slow mode propagation in both the Cartesian slab \citep{edwinslab,solerSF} and the cylinder \citep{edwinroberts}. It is well-known that the slow mode in a $\beta < 1$, homogeneous medium is mainly polarized along the magnetic field direction. So, if one considers a magnetic structure, say, a slab or a cylinder, the slow mode is almost insensible to the perpendicular geometry to magnetic field lines and is mostly governed by the physical conditions internal to the structure. In such a case, the slow mode frequency is well approximated by solving the dispersion relation corresponding to a homogeneous medium with the physical conditions of the internal structure, and selecting an appropriate value for the perpendicular wavenumber to the magnetic field, namely $k_\perp$. This $k_\perp$ contains the effect of the geometry of the magnetic structure. Thus, to apply this technique we consider the dispersion relation for a partially ionized, homogeneous, and infinite plasma derived by \citet[][Equation~(24)]{forteza}, which in our notation is
\begin{eqnarray}
 && \omega^4 - i \left( k_z^2 + k_\perp^2 \right) \eta_{\rm C} \omega^3 - \left( \cs^2 + \va^2 \right) \left( k_z^2 + k_\perp^2 \right) \omega^2 \nonumber \\ 
&&+ i \cs^2 \left( k_z^2 + k_\perp^2 \right) \left[  \left( k_z^2 + k_\perp^2 \right) \eta_{\rm C} -   k_\perp^2 \Xi \rho_0 \va^2 \right] \omega + \left( k_z^2 + k_\perp^2 \right) k_z^2 \cs^2 \va^2 = 0. \label{eq:disperslow}
\end{eqnarray}
Note that in \citet{forteza} the magnetic field is orientated along the $x$-direction, so their $k_x$ and $k_z$ correspond to our $k_z$ and $k_\perp$, respectively. We are interested in approximating the slow mode frequency near the critical $k_z$, where $\omega$ is small and tends to zero. In such a situation, it seems reasonable to neglect the terms with $\omega^3$ and $\omega^4$ in Equation~(\ref{eq:disperslow}). Also, we neglect the term with $\Xi$. So, the dispersion relation becomes a second order polynomial for the frequency that can be solved exactly,
\begin{equation}
 \omega \approx \pm \sqrt{\ct^2 k_z^2 -  \frac{\ct^4 \eta_{\rm C}^2}{4 \va^4}  \left( k_z^2 + k_\perp^2 \right)^2 } + i \frac{\ct^2 \eta_{\rm C}}{2 \va^2}  \left( k_z^2 + k_\perp^2 \right). \label{eq:approxslow}
\end{equation}
We check the validity of this approximation by taking $\eta_{\rm C} = 0$, so $\omega \approx  \pm \ct k_z $, which is consistent with the slow mode frequency in the ideal case. The critical wavenumber, $k_z^{cs}$, can be derived by setting the real part of Equation~(\ref{eq:approxslow}) equal to zero,
\begin{equation}
 k_z^{cs} = \frac{\va^2}{\ct \eta_{\rm C}} \pm \sqrt{\frac{\va^4}{\ct^2 \eta_{\rm C}^2} -k_\perp^2}.\label{eq:critslow}
\end{equation}
In principle, we obtain two critical values from Equation~(\ref{eq:critslow}). However, that given by the $+$ sign is not consistent neither with the small-frequency approximation nor with our numerical results because it corresponds to a very large value of $k_z$. On the contrary, that given by the $-$ sign is the critical value that we are looking for, and for $k_\perp^2 \ct^2 \eta_{\rm C}^2 / \va^4 \ll 1$ it can be approximated by
\begin{equation}
 k_z^{cs} \approx \frac{k_\perp^2 \ct \eta_{\rm C}}{2 \va^2}. \label{eq:critslowfin}
\end{equation}
Hence, the critical wavenumber is both determined by the geometry, through $k_\perp$, and by the ionization degree, through $\eta_{\rm C}$. As a consequence, the slow mode has no critical wavenumber in the unbounded medium, studied by \citet{fortezanonad}. 

To perform a comparison between the analytical approximation (Equation~(\ref{eq:approxslow})) and the numerical results, we have to provide a value for $k_\perp$. The dominant velocity perturbation for the slow mode is $v_z$. Again, we consider the well-known behavior of perturbations in the ideal case and assume that, for $m=1$, $v_z$ is approximately proportional to the Bessel function $J_1$. Let us also assume that the corona has a negligible effect and the slow mode is fully confined within the filament thread. In such a situation, we can approximate $k_\perp^2 \approx j_1^2/a^2 + m^2/a^2$, where the term $j_1^2/a^2$ is the radial contribution to the perpendicular wavenumber, while $m^2/a^2$ represents the azimuthal contribution. We take $\tilde{\mu}_f = 0.8$ and compute the approximate slow mode frequency from Equation~(\ref{eq:approxslow}). This result is plotted by means of symbols in Figure~\ref{fig:slow}. We see that the approximation is reasonably good for $k_z a < 1 $ and near the critical $k_z$, so the approximation describes well the presence of a critical wavenumber for the slow mode. However, it diverges from the numerical result for $k_z a > 1$ because the assumptions considered for deriving Equation~(\ref{eq:approxslow}) are not satisfied for large $k_z a$. 

It is possible to know the relative position of the slow mode critical wavenumber ($k_z^{cs}$  from Equation~(\ref{eq:critslowfin})) with respect to the Alfv\'en mode lower critical wavenumber ($k_z^{c-}$ from Equation~(\ref{eq:alfvencriticsmasmenos})),
\begin{equation}
 k_z^{cs} \approx \frac{\eta_{\rm C}}{\eta} \frac{\ct}{\va} k_z^{c-}.
\end{equation}
For a partially ionized plasma $\eta_{\rm C} \gg \eta$, so we expect $ k_z^{cs} \gg k_z^{c-}$. However, in the fully ionized case, $\eta_{\rm C} = \eta$ but $\ct < \va$, therefore $ k_z^{cs} < k_z^{c-}$. This is in agreement with our results, e.g., compare the lower critical wavenumbers in Figures~\ref{fig:alfven}($a$) and \ref{fig:kink}($a$), on one hand, and Figure~\ref{fig:slow}(a), on the other hand.

\begin{figure}[!ht]
\centering
\epsscale{0.5}
\plotone{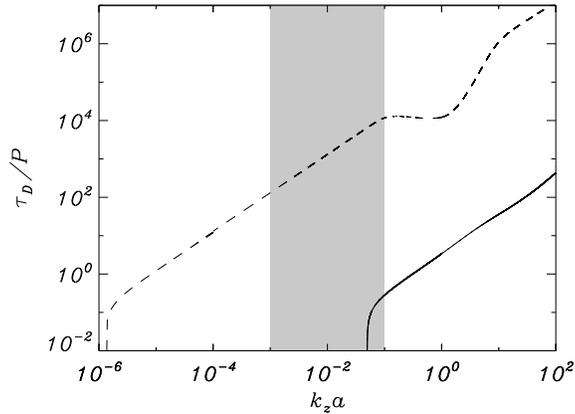}
\caption{Same as Figure~\ref{fig:alfvendamping} but for the slow mode. The dotted and solid lines are superimposed. \label{fig:dampingslow}}
\end{figure}

Finally, the particular effect of ohmic diffusion and ion-neutral collisions on the the slow mode damping is assessed in Figure~\ref{fig:dampingslow}. We obtain that slow mode damping is entirely governed by ion-neutral collisions. The ratio $\tau_{\rm D}/P$ of the slow mode achieves very small values close to the position of the critical wavenumber. This is because the real part of the frequency tends to zero and so the period tends to infinity. It is worth noting that ohmic diffusion never becomes important for the slow mode damping because the presence of the critical wavenumber does not allow the slow mode to propagate for small values of $k_z a$. In addition, Figure~\ref{fig:slow}($b$) indicates that in the case of strong ionization, the ion-neutral collision damping is less efficient than the damping by nonadiabatic effects (thermal conduction and radiative losses) studied in \citet{solernonad}. Only when $\tilde{\mu}_f$ is large, ion-neutral collisions become more efficient than nonadiabatic mechanisms for the slow mode damping.

\section{CONCLUSION}
\label{sec:conclusion}

In this paper, we have studied the propagation of MHD waves in a partially ionized filament thread. Contrary to the fully ionized, ideal case, we have found that wave propagation is constrained by the presence of critical values of the longitudinal wavenumber. This result is of special relevance in the case of the slow magnetoacoustic mode, since the critical wavenumber lies within the observed range of wavelengths of prominence oscillations. This might cause the slow wave not to propagate in thin filament threads and, therefore, it might not be observationally detectable. Turning to the wave damping, we have obtained that both Alfv\'en waves and magnetoacoustic waves are damped by ion-neutral collisions and ohmic diffusion. In the case of Alfv\'en and kink waves, ion-neutral collisions (by means of Cowling's diffusion) dominate for short wavelengths, while ohmic diffusion is more important for large wavelengths. On the contrary, the slow mode is totally governed by Cowling's diffusion. With the exception of the slow mode, theses mechanisms cannot provide damping times compatible with those observed when typical values of the wavelength are considered. In the case of both Alfv\'en and kink waves, one has to consider very short wavelengths and almost neutral plasmas to obtain realistic values of the damping time.
 
One must be aware that in this work we have considered a plasma composed only by hydrogen. So, it is worth to estimate the effect of the presence of helium on the results. Helium remains mostly neutral for typical prominence temperatures, and therefore its presence would increase the value of $\tilde{\mu}_f$ and enhance the effect of ion-neutral collisions. As a consequence, the transitional $k_z a$ at which ion-neutral collisions become important might be shifted toward smaller values, meaning that Cowling's diffusion may dominate for longer wavelengths and/or thinner threads than in a pure hydrogen plasma. Also, the damping time of the three wave modes could be reduced by the presence of helium. On the other hand, since helium only corresponds to 10\% of the prominence material approximately, it could have a minor influence, although its precise effect should be assessed in detail by future investigations.

This paper is a first step in the study of MHD waves in partially ionized filament threads. There are several issues of special interest that may be broached in future works. Some of them are, for example, to assess the effect of neutrals on the resonant absorption of kink waves in the Alfv\'en continuum, to study the spatial damping of waves, and to investigate the time-dependent, initial value problem. 

\acknowledgements{
     The authors acknowledge the financial support received from the Spanish MICINN, FEDER funds, and the Conselleria d'Economia, Hisenda i Innovaci\'o of the CAIB under Grants No. AYA2006-07637 and PCTIB-2005GC3-03. RS thanks the CAIB for a fellowship.}

\appendix

\section{EXPRESSIONS FOR THE PARAMETERS USED IN THIS PAPER}
\label{app:parameters}

We give here expressions for the quantities $\eta$, $\eta_{\rm C}$, and $\alpha_n$ that appear in the text. All quantities are expressed in MKS units. Ohmic and Cowling's diffusivities are given by Coulomb's, $\sigma$, and Cowling's, $\sigma_{\rm C}$ conductivities, respectively,
\begin{equation}
 \eta = \frac{1}{\mu_0 \sigma}, \qquad  \eta_{\rm C} = \frac{1}{\mu_0 \sigma_{\rm C}}, 
\end{equation}
with
\begin{equation}
 \sigma = \frac{n_e e^2}{m_e \left( \nu'_{ei} + \nu'_{en} \right)}, \qquad \sigma_{\rm C} = \frac{\sigma}{1+\xi_n^2 B_0^2 \sigma / \alpha_n},
\end{equation}
where $m_e$ and $e$ are the electron mass and charge, and $\alpha_n$ is the friction coefficient, given by 
\begin{equation}
 \alpha_n = m_e n_e  \nu'_{en} + m_i n_i  \nu'_{in}.
\end{equation}
There $\nu'_{ei}$, $\nu'_{en}$, and $\nu'_{in}$ are the effective electron-ion, electron-neutral, and ion-neutral collisional frequencies, respectively, defined as
\begin{equation}
 \nu'_{ei} = \frac{m_i}{m_i + m_e} \nu_{ei},  \qquad  \nu'_{en} = \frac{m_n}{m_n + m_e} \nu_{en}, \qquad \nu'_{in} = \frac{m_n}{m_n + m_i} \nu_{in},
\end{equation}
with
\begin{equation}
 \nu_{ei} = 3.7 \times 10^{-6} \frac{n_i \Lambda Z^2}{T_0^{3/2}}, \qquad \nu_{en} = n_n \sqrt{\frac{8 k_{\rm B} T_0}{\pi m_{en}}} \Sigma_{en}, \qquad \nu_{in} =  n_n \sqrt{\frac{8 k_{\rm B} T_0}{\pi m_{in}}} \Sigma_{in},
\end{equation}
where $\Sigma_{in} = 5 \times 10^{-19}$~m$^2$ and $\Sigma_{en} = 10^{-19}$~m$^2$ are the ion-neutral and electron-neutral cross-sections, respectively, $\Lambda$ is de Coulomb logarithm, and
\begin{equation}
 m_{in} = \frac{m_i m_n}{m_i + m_n}, \qquad  m_{en} = \frac{m_e m_n}{m_e + m_n}.
\end{equation}
For a hydrogen plasma, $Z=1$ and $m_i \approx m_n$, so a simplified expression for the friction coefficient $\alpha_n$ can be provided by neglecting the contribution of electron-neutral collisions,
\begin{equation}
 \alpha_n = \frac{1}{2} \xi_n \left( 1 - \xi_n \right) \frac{\rho_0}{m_n} \sqrt{\frac{16 k_{\rm B} T_0}{\pi m_i}} \Sigma_{in}.
\end{equation}
In a fully ionized plasma, $\xi_n = 0$, hence $\sigma_{\rm C} = \sigma$ and $\alpha_n = 0$. Also note that in a fully neutral plasma,  $\xi_n = 1$ and $n_i = n_e = 0$, so $\sigma = \sigma_{\rm C} = \alpha_n = 0$.

\end{document}